\documentclass[%
 reprint,
superscriptaddress,
 amsmath,amssymb,
 aps,
]{revtex4-2}

\usepackage{graphicx}
\usepackage{dcolumn}
\usepackage{bm}

\begin{document}


\title{Quantum Networks for High Energy Physics (HEP)}


\author{Andrei Derevianko}
\affiliation{Department of Physics, University of Nevada, Reno, 89557, USA}

\author{Eden Figueroa}
\affiliation{Physics and Astronomy Department, Stony Brook University}
\affiliation{Quantum Information Science \& Technology Laboratory, Instrumentation Divison,  Brookhaven National Laboratory, Upton NY 11973}

\author{Juli\'{a}n Mart\'{i}nez-Rinc\'{o}n}
\affiliation{Quantum Information Science \& Technology Laboratory, Instrumentation Divison,  Brookhaven National Laboratory, Upton NY 11973}

\author{Inder Monga}
\affiliation{Scientific Networking Division, Lawrence Berkeley National Laboratory}

\author{Andrei  Nomerotski}
\affiliation{Physics Department, Brookhaven National Laboratory, Upton NY 11973}

\author{Cristi\'{a}n H. Pe\~na}
\affiliation{Fermi National Accelerator Laboratory, Batavia, IL 60510}

\author{Nicholas A. Peters}
\affiliation{Computational Sciences and Engineering Division, Oak Ridge National Laboratory}

\author{Raphael  Pooser}
\affiliation{Computational Sciences and Engineering Division, Oak Ridge National Laboratory}

\author{Nageswara Rao}
\affiliation{Computational Sciences and Engineering Division, Oak Ridge National Laboratory}

\author{Anze Slosar}
\affiliation{Physics Department, Brookhaven National Laboratory, Upton NY 11973}

\author{Panagiotis Spentzouris} 
\affiliation{Fermi National Accelerator Laboratory, Batavia, IL 60510}

\author{Maria Spiropulu}
\affiliation{Division of Physics, Mathematics and Astronomy, California Institute of Technology, 1200 E California Blvd., Pasadena, CA 91125}

\author{Paul Stankus}
\affiliation{Quantum Information Science \& Technology Laboratory, Instrumentation Divison,  Brookhaven National Laboratory, Upton NY 11973}

\author{Wenji Wu}
\affiliation{Scientific Networking Division, Lawrence Berkeley National Laboratory}

\author{Si Xie}
\affiliation{Fermi National Accelerator Laboratory, Batavia, IL 60510}
\affiliation{Division of Physics, Mathematics and Astronomy, California Institute of Technology, 1200 E California Blvd., Pasadena, CA 91125}

\date{\today}

\begin{abstract}
Quantum networks of quantum objects promise to be exponentially more powerful than the objects considered independently.  To live up to this promise will require the development of error mitigation and correction strategies to preserve quantum information as it is initialized, stored, transported, utilized, and measured. The quantum information could be encoded in discrete variables such as qubits, in continuous variables, or anything in-between.   Quantum computational networks promise to enable simulation of physical phenomena of interest to the HEP community.  Quantum sensor networks promise new measurement capability to test for new physics and improve upon existing measurements of fundamental constants.    Such networks could exist at multiple scales from the nano-scale to a global-scale quantum network.\footnote{This manuscript has been authored by UT-Battelle, LLC under Contract No. DE-AC05-00OR22725 with the U.S. Department of Energy. The United States Government retains and the publisher, by accepting the article for publication, acknowledges that the United States Government retains a non-exclusive, paid-up, irrevocable, world-wide license to publish or reproduce the published form of this manuscript, or allow others to do so, for United States Government purposes. The Department of Energy will provide public access to these results of federally sponsored research in accordance with the DOE Public Access Plan (http://energy.gov/downloads/doepublic-access-plan).}

\end{abstract}

\maketitle


\section{Introduction}
Quantum technologies manipulate individual quantum states and make use of superposition, entanglement, squeezing, and backaction evasion. Quantum sensors \cite{quantumsensing, Gorshkov2018} exploit these phenomena to make measurements with a precision better than the Standard Quantum Limit (SQL), with the ultimate goal of reaching the Heisenberg Limit. A single quantum sensor can only take advantage of quantum correlations in a single location, while a quantum network could exploit the correlations across an array of sensors, linking them to each other with quantum mechanical means. This improves the sensitivity and scalability of the resulting entangled system simultaneously allowing it to benefit from the long-distance baseline between the sensors. In Section II below we list several options for possible quantum networks with quantum sensors and give examples of how these systems can be used for the HEP science. Then in Section III we briefly review current status and challenges for quantum networks.

\section{HEP topics benefiting from development of quantum networks - existing ideas and protocols}

Below we discuss several ideas how the HEP measurements can benefit from quantum networks.

\subsection{Quantum-assisted telescopes}

High precision astrometry at the microarcsecond level potentially can be achieved with optical interferometers connected in a quantum network. It could open science avenues for imaging black hole accretion disks, improving the local distance ladder, detailing dark matter subhalo influence on microlensing, and dark matter impact visible in Galactic stellar velocity maps \cite{astrometryWP}. This could be enabled by new ideas crosscutting optical interferometry and quantum information science.

Observations using interferometers provide sensitivity to features of images on angular scales much smaller than any single telescope.  Traditional (Michelson stellar) optical interferometers are essentially classical, interfering single photons with themselves \cite{Pedretti2009, Martinod2018, tenBrummelaar2005}, and the single-photon technique is highly developed and approaching technical limits.  Qualitatively new avenues for optical interferometery can be opened up, however, once we consider using multiple-photon states; these generally require a quantum description, especially in conjunction with non-classical quantum technologies such as single-photon sources, entangled pair sources, and quantum memories.  We will focus here on a particular two-photon state technique with application for precision astrometry.

It has been recently proposed that stations in optical interferometers would not require a phase-stable optical link if instead sources of quantum-mechanically entangled pairs could be provided to them, potentially enabling hitherto prohibitively long baselines \cite{Gottesman2012}. If these entangled states could then be interfered locally at each station with an astronomical photon that has impinged on both stations, the single photon counts at the two stations would be correlated in a way that is sensitive to the phase difference in the two paths of the photon, thus reproducing the action of an interferometer.

Several variations of this idea have been proposed. For one of them, which perhaps is a longer term for practical implementation, high intensity wide-band sources of entangled photons and quantum memories would be employed to measure correlations between the stations as explained above \cite{harvard1}. The approach can be generalized from the entanglement of photon pairs to multipartite entanglement in multiple stations and quantum protocols to process information in noisy environments for evaluation of experimental observables. In another approach, which potentially could be implemented in a shorter term, two photons from different sky sources are interfered at two separate stations, requiring only a slow classical information link between them \cite{stankus2020}. The latter scheme can be contrasted with the Hanbury Brown \& Twiss intensity interferometry \cite{hbt} and could allow robust high-precision measurements of the relative astrometry of the two sources. A calculation based on photon statistics suggests that angular precision on the order of $10\mu$as could be achieved in a single night’s observation of two bright stars \cite{stankus2020}. 

Increased sensitivity to fainter objects can be achieved for the schemes with bright entangled photon sources and quantum memories \cite{harvard2} employing technologies, which are under development for quantum networks. Though it looks quite futuristic now, the field of quantum information science is going through exponential expansion driven by the industry and, within a decade, may offer capabilities matching the requirements.

An important consideration from the instrumentation viewpoint is that the photons must be close enough in time and in frequency to efficiently interfere; or, formulating it differently, to be indistinguishable within $ \Delta t \cdot \Delta E \sim \hbar $. Converting energy to wavelength, the above is satisfied for $\Delta t \cdot \Delta \lambda = 10~\mathrm{ps} \cdot 0.2~\mathrm{nm}$ at 800~nm wavelength, setting useful target goals for the temporal and spectral resolutions \cite{Nomerotski2020_1}. Another important parameter for the imaging system is the photon detection efficiency, which needs to be as high as possible, since the two-photon coincidences  have a quadratic dependence on it.

An efficient scheme of spectroscopic binning can be implemented by employing a traditional diffraction grating spectrometer where incoming photons are passed though a slit, dispersed, and then focused onto a linear array of single-photon detectors \cite{Dey2019, Vogt1994, Zhang2020}. However, improvement of timing resolution appears to be the most straightforward way to achieve the targeted performance. Fast technologies, such as superconducting nanowire single photon detectors (SNSPD) and single photon avalanche devices (SPAD), can be considered for this application. The superconducting nanowire detectors have excellent photon detection efficiency, in excess of 90\% \cite{Zhu2020, Divochiy2008}, with demonstrated 3~ps timing resolution for single devices~\cite{Korzh2020}.
The SPAD sensors based on silicon avalanche diodes produce fast pulses of big enough amplitude for single photon detection. These devices also have excellent timing resolution, which can be as good as 10~ps for single-channel devices, and most importantly, good potential for scalability with multi-channel imagers already reported \cite{Gasparini2017, Morimoto2020}. Benchmarking these promising technologies for a spectrograph with required spectral and timing resolutions is currently in progress~\cite{nomerotski2021}.

\subsection{Atomic clocks in a quantum network}


Atomic clocks are arguably the most accurate quantum sensors ever built, with modern clocks guaranteed to be accurate to a fraction of a second over the age of the Universe. Developed over many decades, the atomic clock toolbox forms a natural platform for quantum information processing with neutral atoms and ions. In atomic clocks, an atomic transition serves as a frequency reference for an external source of electromagnetic radiation, microwave or optical cavity. The frequency source is tunable and once its frequency is in resonance with the atomic transition, the period of oscillation is fixed and one counts the number of oscillations at the source. Then the measured time is simply the number of counted oscillations $\times$ known oscillation period determined by the atomic transition frequency.  

In atomic clocks, the quantum oscillator is well-protected from the environment and  the conventional Standard Model physics is well-controlled and characterized. This opens up intriguing prospects for exotic physics searches. One of the key ideas in this context  is that the clocks are sensitive to  variations of fundamental constants, such as the fine-structure constant $\alpha$. While the clocks have been historically employed in sensitive searches for slow drifts of fundamental constants~\cite{RosHumSch08,Huntemann2014}, more recent ideas involve a broad range of regimes of how the constants can vary: transient~\cite{DerPos14}, oscillating~\cite{ArvHuaTil15}, and stochastic~\cite{Derevianko2016a} regimes. 

Atomic clocks have enjoyed rapid progress. The advent of frequency combs led to a revolution in the field of frequency metrology \cite{fortier201920}. It stimulated the development of optical atomic clocks. A few optical clocks operated with single ions feature a record control of the systematics in the 10\textsuperscript{-17} range. The next big leap for atomic clock  is optical lattice clocks  \cite{DerKat11,bloom2014optical}. For an optical lattice clock, thousands of probed atoms allow unprecedented statistical resolutions that translate into frequency stabilities of a few 10\textsuperscript{-18} at one second. The key feature of these types of clocks is the trapping of neutral atoms in a deep optical potential, which allows interrogation times of several 100~ms. Optical lattice clocks containing many atoms have demonstrated stability that reaches the standard quantum limit. 

Recent research in quantum atomic clocks \cite{Kmr2014,komar2016quantum} showed that a quantum network of atomic clocks can result in a substantial boost of the overall precision if multiple clocks are phase locked and connected by quantum entanglement. Entanglement creates nonlocal quantum correlation among remote atoms. If nonlocal quantum correlation is properly set up and employed in the optimal way, significant noise reduction can be achieved. Compared to a single clock, the ultimate precision will improve as 1/K, where K is the number of clocks. If the same clocks are connected via a classical network, the precision scales as $1/\sqrt{K}$. Ultimately, a quantum network of atomic clocks can surpass the SQL to reach the ultimate precision allowed by quantum theory — the Heisenberg limit. 

Today, the world’s most accurate atomic clock has reached a frequency accuracy of 10\textsuperscript{-21}\cite{bothwell2022resolving}. Higher accuracy atomic clocks of 10\textsuperscript{-22} and beyond potentially can be achieved by entangling their atoms in a quantum network, which has the potential to transform global timekeeping, enabling orders-of-magnitude improvements in measurement accuracy and sensor resolution for a wide range of scientific and technological applications. They also have strong impact on many fundamental research areas, including  searches for dark matter and dark energy, gravitational physics, and quantum many-body physics.  The challenge in realizing a quantum network of atomic clocks is that the entangled state needs to be generated with both high fidelity and high rate (to maximise the measurement duty cycle) over long distance.

Geographically-distributed networks are paramount in searches for dark matter transients~\cite{DerPos14} and exotic field bursts from powerful astrophysical events such as black-hole mergers~\cite{dailey2020ELF.Concept}. A network enables tracking the sweep of the nodes by the transient yielding the directionality and velocity of the event. Moreover, individual network nodes have to be affected by the transient in a certain time pattern, providing a crucial vetoing mechanism. First results for dark-matter wall searches with atomic clock networks were reported using the archival data from $\sim 30$ microwave atomic clocks onboard GPS satellites~\cite{Roberts2017-GPS-DM} and a trans-European network of optical clocks~\cite{Roberts2019-DM.EuropeanClockNetwork}. As of now, there are no searches  with a cross-node entangled network of atomic clocks, as such networks are in their infancy~\cite{Nichol2021}.

The utility of entanglement in the searches for transients has been discussed in Ref.~\cite{Daykin2021}: For individual sensors, the entanglement or spin-squeezing is a useful resource, as it improves the sensitivity of a single-shot measurement typical of a search for short transients. Cross-node or geographically-distributed entanglement~\cite{komar2016quantum} is not useful and is, in fact,  detrimental to the network searches for transients. Indeed, a projective measurement on a single node collapses the distributed wave-function, effectively rendering all other nodes deaf to the transient. Then the network loses both velocity and angular resolution.

However, the entangled networks can be useful if all the nodes are affected by an exotic field that is nearly uniform across the network. In the context of dark matter candidates such as dilatons and moduli (these lead to oscillating fundamental constants), such conditions are realized when the correlation length of the dark matter field is much larger than the network spatial extent~\cite{Derevianko2016a}.

\subsection{Quantum network for axion searches}

Dark matter may form macroscopic objects, a possibility that arises naturally in theories with topological defects formed in the early Universe. Having a quantum network of sensors would be advantageous because: (i) greatly separated nodes (sensors) have access to spatial information that is unavailable when each node is used independently, which may be exploited to look for long-distance effects of new physics; (ii) each node in the network can be precisely monitored by optical signals, so they may serve as real-time particle-effect detectors. The multiplicity of the nodes would enhance the detection capabilities by increasing the number of scattering targets; and very importantly, (iii) the entanglement of the network allows the use of quantum metrology techniques to increase the sensitivity and bandwidth of the experiments.

A quantum network of atomic ensembles may be reconfigured as a network of entangled atomic clocks by making use of the existing quantum memories and augmenting them with interferometers at each node. Spatially extended dark matter may couple to the Standard Model and induce transient changes in fundamental constants, such as the fine structure constant. An encounter of the spatially extended dark matter with one or more nodes would then lead to a transient shift in the energy level splittings of the atomic systems, desynchronizing the corresponding atomic clock node(s) with respect to the rest of the network.

Dark matter could be detected by measuring the clock phase correlations across the network. Since the signal relies on measuring desynchronization between clocks, it is crucial that the atomic clocks are spatially separated, so that the dark matter does not interact with all clocks at the same time as it moves across the network. Potentially, this quantum network of atomic clocks could be implemented with faster timing synchronization between clocks and higher clock precision as compared with atomic clocks used in the GPS system. In this, it is important to achieve synchronization of few nanoseconds in a tens or hundreds km long fiber network. Such capabilities may lead to an increase in the sensitivity over existing GPS-based searches for spatially extended dark matter~\cite{Cline-2007} in significant parts of parameter space.

Axionlike, scalar, and millicharged particles are well-motivated possibilities for new physics beyond the Standard Model, and they may also be all or a component of dark matter. Millicharged particles would have small couplings with photons, and many models also predict small interactions between axionlike or scalar particles and photons. These particles may be detected through a sensitive measurement of light polarization or phase. Such light-polarimetry experiments have already been performed to look for axionlike, scalar and millicharged particles at Brookhaven National Laboratory and other institutions~\cite{NA64-2020}. In these experiments, a beam of polarized light is directed towards a magnetic field region. In the magnetic field, photons may mix with axionlike or scalar particles, or occasionally pair-produce milli-charged particles. A polarization measurement can then be performed to detect changes in the net polarization of the photon beam caused by these new particles. This effect is enhanced with the distance travelled by the beam. Current and past experiments have considered a beam travelling over a few meters.

It is needed to investigate how these experiments can be scaled up by a quantum network and to calculate the resulting sensitivity to dark-sector particles. In this setup, two or more photon beams with orthogonal polarizations, or pairs of polarization-entangled photons, may be distributed across the network nodes. Light beams with polarization that is parallel to the Earth’s magnetic field, or to an artificially generated external magnetic field, would experience a small rotation in their polarization via interactions with axions or millicharged particles. In particular, for axionlike and scalar particles, the effect on the change in polarization grows quadratically with the node separation. This experimental concept would thus significantly benefit from a long baseline. A measurement of the polarization of different beams across the network may then be performed to look for these dark-sector particles. Ultimately, such an experiment should be performed in vacuum, limiting it to a medium size network.

\subsection{Magnetometers in a quantum network} 
Ultra-sensitive detection of magnetic fields has many applications, including the detection of WIMPs \cite{xu2006magnetic,xia2006magnetoencephalography,bick1999hts,dougherty2006identification}. Superconducting quantum interference device (SQUID) magnetometers have traditionally maintained the highest sensitivities, but recently atomic magnetometry has reached parity, with sensitivity below 1~fT/$\sqrt{Hz}$~\cite{Kominis:2003,Budker:2007}. Sensitivity in these systems is limited by atomic projection noise and photon shot noise~\cite{PhysRevA.62.043403,PhysRevLett.95.063004}. Experimental sources have demonstrated both back-action evasion and quantum noise reduction of shot noise~\cite{PhysRevLett.104.133601,PhysRevLett.105.053601,PhysRevA.86.023803}. Multiple demonstrations have shown that it is possible to squeeze the optical readout field in these devices~\cite{PhysRevLett.105.053601,PhysRevA.86.023803}, including performing two-mode squeezing operations and magnetic field transduction in the same medium~\cite{otterstrom_nonlinear_2014}. 
The sensitivity of an atomic magnetometer measurement can be described by the sampling of an ensemble of N polarized atoms over a measurement time t: 
$
\delta B_{min}=\frac{1}{\gamma_{e}\sqrt{NTVt}}, \label{b_min}
$
where T represents the coherence lifetime, $\gamma_{e}$ the atomic gyromagnetic ratio and V is the cell volume.  

Applications of atomic magnetometers have thus far been developed for measurements of DC, RF, and gradient magnetic fields. While their overall components are similar, they differ in their operation.
Historically, DC or low-frequency AC magnetometers have been designed to operate at relativity low vapor pressures to avoid spin-exchange interactions, which limit coherence time and sensitivity.  
The development of spin-exchange relaxation free (SERF) devices has brought the promise of sensitivity approaching atto-Tesla~\cite{Kominis:2003}, while increased coherence is obtained via vapor cell coatings~\cite{Seltzer:2009}.  
Detection of RF fields in the range of kHz to MHz has also been demonstrated \cite{Savukov:2005}, and magnetic field gradient measurements have been obtained with high buffer gas configurations and pixelated detectors~\cite{Kominis:2003}. 
Notably, miniaturized devices have been demonstrated~\cite{Kitching:2018} to reduce the size, weight, and power requirements while still maintaining state of the art sensitivity. 

\subsection{Gamma Spectroscopy: Remote Quantum Computing}

We speculate that gamma spectrum sensors connected to remote quantum computers
via a Quantum Internet (QI) may provide input to quantum computations that directly operate on the spectra, possibly in adiabatic mode. This is a similar concept to coupling scintillation light into optical fiber to get it out of a high magnetic field environment for analysis~\cite{HAAK1997455}. 
Such networking capability has a potential to provide unprecedented computing precision in a number of HEP scenarios, particularly in nuclear areas. Gamma sensing is often motivated by nuclear non-proliferation and security scenarios that involve inferences about facility operations and material movements.

Several nuclear isotopes of interest generate spectral signatures in 3 MeV range (for example, 661 KeV of Cs-138) and NaI and HPGe detectors are commonly used for obtaining their Gamma spectra using photo multipliers.
In practice, these spectral measurements are used to generate discretizations that are subsequently processed digitally and transferred over conventional networks. 
This approach is inherently subject to discretization errors, which persist when they are given as input to quantum processing algorithms that operate on the Gamma spectra.
The ability to transport photo-multiplier photons to remote quantum computations avoids the inherent quantization limits, for example, that may result in missing of peaks in very narrow spectral regions. 

Photons from photo multipliers in Gamma spectra are typically in 1~nm range, and are suitable for network transport that operates in C-band, for example, 1550~nm wavelength.
These photons need to to be transducted into C-band for QI communications.
The transduction devices may be attached to photo multipliers at the sensor end and QI on the other end, and for reverse purpose at the remote computing systems.
By book-ending a QI connection with pair of transduction devices, the photo-multiplier output can be faithfully replicated and provided as input to quantum computations by avoiding the current discretization errors.

\subsection{Quantum Networked accelerometers}
Accelerometers have been proposed as a mechanism to probe the coupling of dark matter to test masses via gravity~\cite{Carney2021}.  It is argued that a wide range of potential mechanical sensing technologies opens the possibility of searching for dark mater over a wide range of energy scales and coupling mechanisms~\cite{Carney2021}.

\section{Current status and challenges for the development of quantum repeater networks}

For the networked applications described in the previous section, the distribution of entanglement is a crucial resource. Presently, spontaneous parametric down-conversion (SPDC) has become the most common source of entanglement for quantum networks. In SPDC, a strong classical laser pumps a nonlinear crystal, resulting in pairs of photons. For the applications mentioned above, photons from different sources must fulfill strict requirements, including indistinguishability. This requirement is especially important for protocols that rely on two-photon Hong-Ou-Mandel quantum interference, which is at the heart of Bell state analysis used for quantum teleportation. 

The simplest distribution of entanglement between a fixed pair of destinations uses a source of entangled pairs, which then are sent to the two locations. This transmission from a single source ultimately will be limited in distance by the attenuation in the communication channels. Practically, the attenuation of single photons in present commercial telecommunications fibers will limit single fiber transmission to distances and there is a fundamental quantum communications rate-loss trade off bound for repeater-less channels~\cite{PLOB}.  So, for example, after transmission through 100 km of standard single mode optical fiber, there will be 20 dB of attenuation on a transmitted photon, if two entangled photons are each transmitted over 100 km, then the throughput would be reduced by 40 dB.  As a result, quantum repeaters are required to build large-scale quantum networks with high throughput. 

One of the original ideas to make a quantum repeater relies on breaking a two fiber link into four shorter segments and placing entangled photon pair sources in the middle of where the first link was split. The quantum repeater can “hop” the entanglement property across the larger distance interval by consuming the resource of a second entangled pair. The two innovations needed to carry this out are the quantum process of entanglement swapping and the existence of quantum memories. In this approach to a quantum repeater, two sources generate entangled pairs, with the pairs initially independent of each other. One member of each pair are brought together and interfered at an intermediate location (where the source would be initially if it were a symmetric link), projecting this pair onto one of four Bell states. This partial collapse of the four-particle wavefunction results in the two remaining particles, being in an entangled state—even though they were initially independent. This result is an example of entanglement swapping~\cite{Pan98}. 

The real potential for improved distance transmission with the quantum repeater is realized with the addition of working quantum memories, which can buffer a pair after a successful transmission without requiring the other pair’s survival at the same time. Quantum memories (or buffers) are devices where non-classical states of photons (e.g., single photons, entanglement, or squeezing) are mapped onto stationary matter states and preserved for subsequent retrieval. These devices require sophisticated optical control, enabling coherent interactions at the quantum level between light and matter. The ability of quantum memories to synchronize, buffer, and herald probabilistic events makes them a key component in advanced quantum communications networks.

For application in quantum repeater networks, optical quantum memories can be categorized into two classes. The Class 1 (also called “quantum registers”) emits one photon first and discharges another photon later on demand (refer to the Duan-Lukin-Cirac-Zoller (DLCZ) protocol~\cite{DLCZ}). Class 2 is the in-and-out quantum memory, which can store an arbitrary state of a single photon and release it later on demand. Examples of these memories are the working prototypes based on electromagnetically induced transparency (EIT) or atomic frequency combs (AFCs). Depending on the quantum memories being used, different types of quantum repeaters can be built.  We discuss some major approaches below.

\noindent \textbf{DLCZ Quantum Repeater}. This is experimentally accessible by using Class 1 memories that emit one photon first and another photon later on demand, relying on quantum interference to generate entanglement. A drawback of this approach is its probabilistic nature, which severely limits the entanglement creation rate. Several examples of these repeaters are currently being built.

\noindent \textbf{Entangled Photon Pair Quantum Repeater}. This alternative scheme relies on using entangled photon pair sources interfaced with Class 2 in-and-out quantum memories, capable of receiving and storing an arbitrary state of a single photon and releasing it later on demand. Through heralding on one of the two photons, this approach could be deterministic, thus generating higher entanglement distribution rates. Recently, advanced experiments have performed entanglement swapping with quantum memories.

Current quantum memories are not quantum error corrected, so for these first two types of quantum repeaters, one needs to perform an entanglement purification protocol which consumes multiple low quality entangled pairs to yield a single high quality entangled pair~\cite{Bennett1996}. These protocols require two-way classical communications, so as the network gets larger, the quantum memories generally need to have lower decoherence and longer storage times. 

\noindent \textbf{Quantum Error Corrected Quantum Repeater.}
This type of repeater encodes quantum information or distributes entanglement in multiple photons or photonic modes to form a quantum error correcting code.  The photons are then transmitted over the link and error corrected.  It has been theoretically shown that loss and other operational errors can be corrected to enable continental-scale fault tolerant quantum communications~\cite{PhysRevLett.112.250501}.  Another proposed variation of an error corrected repeater is all-optical~\cite{ATL2015}.  

Building the infrastructure for a first prototype of a quantum repeater still requires large investments to further develop the needed quantum network hardware, including: 

$\bullet$ Quantum-limited detectors, ultra-low loss optical channels, and classical networking and cybersecurity protocols.

$\bullet$ Entanglement sources and transmission, control, and measurement of quantum states.

$\bullet$ Transducers for quantum sources and signals from optical to telecom regimes.

$\bullet$ Development of quantum memory buffers that are compatible with photon-based quantum bits in optical or telecom wavelengths.

$\bullet$ The experimental development of quantum error correction and/or entanglement purification to make up for decoherence errors due to quantum device imperfections.

\section{Conclusions}

Above, we described several ideas how quantum networks can benefit HEP science. We also briefly reviewed the current status and roadmap for quantum networks over the next ten years.  Key research priorities include the development of fundamental building blocks for quantum networks, their assembly into larger networking systems, and eventually demonstrating that fault-tolerant quantum communications may be performed.

Evolution of these ideas will depend very much on theoretical and experimental efforts to verify them, to perform benchmarking measurements, to develop concepts of experiments and to evaluate their sensitivity. As these techniques mature over time, they should be included in the DOE planning process for new experiments.

\section*{Acknowledgements}
This work was performed in part at Oak Ridge National Laboratory (ORNL), operated by UT-Battelle for the U.S. Department of Energy under contract no.  DE-AC05-00OR22725. ORNL funding was provided by the U.S. Department of Energy, Office of Science, Office of Advanced Scientific Computing Research,through the Transparent Optical Quantum Networks for Distributed Science Program and the Office of High Energy Physics Quantised Program (Field Work Proposals ERKJ355 and RKAP63). This work was supported in part by the U.S. National Science Foundation grant PHY-1912465.

%

\end{document}